\documentclass[twocolumn,preprintnumbers, aps, amsmath, amssymb, pra ]{revtex4-2}

\usepackage{float}
\usepackage{latexsym,amssymb,amsthm,amsmath,epsfig}
\usepackage{graphicx}
\usepackage{grffile}
\usepackage{dcolumn}
\usepackage{bm}
\usepackage{hyperref}
\hypersetup{
    citecolor=red,
    colorlinks=true,
    linkcolor=blue,
    filecolor=blue,      
    urlcolor=blue,
}

\usepackage[mathlines]{lineno}

\begin{document}


\title{Zero photon catalysis involved eight state discrete modulated measurement-device-independent continuous-variable quantum key distribution}

\author{Muhammad Bilal Khan$^{1}$}
\author{Muhammad Waseem$^{1,2}$}%

\author{Muhammad Irfan$^{1,2}$}%

\author{Asad Mehmood$^{1}$}%

\author{Shahid Qamar$^{1,2}$}%
\affiliation{
$^{1}$  Department of Physics and Applied Mathematics - Pakistan Institute of Engineering and Applied Sciences (PIEAS), Nilore, Islamabad $45650$, Pakistan}
\affiliation{
$^{2}$ Center for Mathematical Sciences - PIEAS, Nilore, Islamabad $45650$, Pakistan}

\date{\today}

\begin{abstract}
Zero photon catalysis (ZPC) introduces noiseless attenuation and can be implemented by existing technologies in quantum key distribution protocols. In this paper, we present a zero photon catalysis-based eight-state measurement-device-independent continuous-variable quantum key distribution (MDI-CV-QKD) combined with discrete modulation and reverse reconciliation. This ZPC involved eight state protocol shows better efficiency in terms of optimal modulation variances, key rates, transmission distances, tolerable excess noises, and reconciliation efficiency as compared to the eight-state protocol without ZPC, four-state without ZPC, and four-state with ZPC, at low signal-to-noise ratio.
\end{abstract}

\maketitle

\section{Introduction}
\label{intro}
In quantum information processing, quantum key distribution (QKD)~\cite{gisin_quantum_2002} aims to share secrete keys between two distant users Alice and Bob through insecure quantum and classical channels, which are controlled by Eve. One category of QKD protocols is discrete variable (DV) QKD~\cite{BEN84,ekert_quantum_1991, bennett_experimental_1992}.
DV-QKD systems~\cite{BEN84, gessner_efficient_2016, scarani_quantum_2008} have received significant attention in the last three decades and are even commercially available. DV-QKD protocols require a single-photon source and detection~\cite{BEN84}. This requirement may make the DV-QKD protocols superior in terms of the transmission distance but inferior in key rate generations \cite{wang_high_2018, huang_continuous-variable_2015}.

To overcome this shortcoming of the low key rate, a second category is continuous variables (CV) QKD~\cite{ralph_continuous_1999, grosshans_continuous_2002, grosshans_quantum_2003,weedbrook_gaussian_2012, ralph_security_2000, lode,jouguet_experimental_2013,weed, jouguet_preventing_2013,huang_field_2016,huang_long-distance_2016,guo_continuous-variable_2019,zub}.
The CV-QKD protocols utilize homodyne or heterodyne detection rather than photon counters, which makes it practically more effective and compatible with existing optical communication systems~\cite{ralph_security_2000, grosshans_2005, kimble_quantum_2008}.
Especially, Gaussian modulated CV-QKD protocol using coherent states~\cite{grosshans_continuous_2002} has been proved to be secure under arbitrary collective attacks~\cite{navascues_optimality_2006}, coherent attacks~\cite{renner_finetti_2009}, and even taking into account the finite size regime~\cite{leverrier_security_2013,leverrier_security_2017, leverrier_composable_2015}.
Furthermore, Gaussian modulated CV-QKD protocols have been implemented in laboratory~\cite{grosshans_quantum_2003,jouguet_experimental_2013,huang_field_2016, huang_long-distance_2016, zhang_continuous-variable_2019}.
However, the traditional CV-QKD system lacks long-distance communications as compared with DV-QKD. To distribute the secret key over a longer distance, an unconditionally secure CV-QKD protocol with discrete modulation has been introduced~\cite{leverrier_unconditional_2009}.

In a practical scenario, imperfections in the devices lead to insecure loopholes, which Eve can exploit to perform quantum attacks. These attacks include local oscillator fluctuation attack~\cite{ma_local_2013}, calibration attack~\cite{jouguet_preventing_2013}, local oscillator wavelength attack~\cite{ma_wavelength_2013}, and detector saturation attack~\cite{qin_quantum_2016, huang_quantum_2013}.
Usually, there are two approaches to removing all these security loopholes. One is device-independent QKD~\cite{acin_device-independent_2007, marshall_device-independent_2014} based on a violation of Bell's inequality, and has been demonstrated in experiments very recently~\cite{nadlinger_experimental_2022, zhang_device-independent_2022,liu_toward_2022}. 
The second one is measurement-device-independent (MDI)-CV-QKD~\cite{pirandola_high-rate_2015, zhang_finite-size_2017, li_continuous-variable_2014, ma_gaussian-modulated_2014, braunstein_side-channel-free_2012, lo_measurement-device-independent_2012, ottaviani_continuous-variable_2015, papanastasiou_finite-size_2017, chen_composable_2018}, which provides a more practical way to prevent all side-channel attacks on detection~\cite{papanastasiou_finite-size_2017}.

\begin{figure*}[t]
\includegraphics[width=6.0 in]{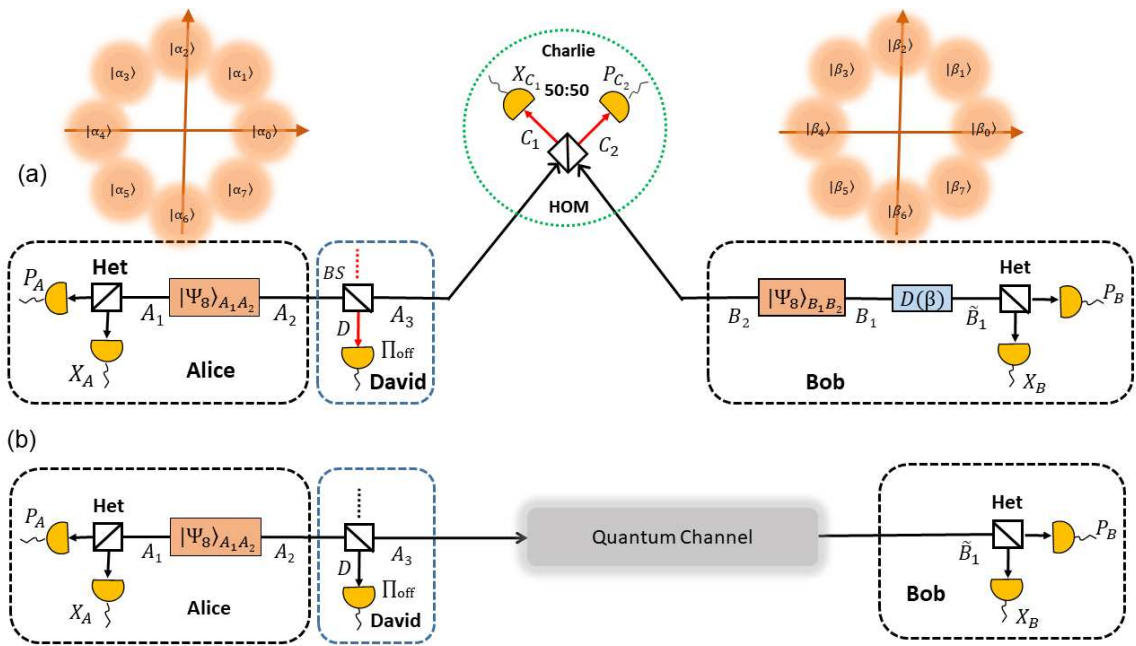}
\caption{(a) Schematics of EB version of MDI-CV-QKD with discrete modulation involving ZPC operation performed by David.
(b) Equivalent one way MDI-CV-QKD protocol with ZPC operation assuming that Bob prepared state $\vert\mathrm{\Psi_{8}}\rangle_{B_{1}B_{2}}$ and displacement are untrusted except for heterodyne detection. Further details are given in the text.}
\label{fig:1}
\end{figure*}

The security of the MDI-CV-QKD protocol does not depend on the reliability of measurement devices.
In MDI-CV-QKD, Charlie is introduced between both Alice and Bob~\cite{pirandola_high-rate_2015}. Charlie performs the Bell state measurement on the states sent by both Alice and Bob and then he publicly announces his measurement results.
Alice and Bob perform the post-processing steps to generate the key rates. 
As the measurement is performed by a third party, the security of the protocol no longer depends on device perfections.
Therefore, it is possible to remove all known and unknown side-channel attacks on detectors in MDI-CV-QKD protocols, which results in higher practical security.
In this regard, several significant results have been achieved in different theoretical~\cite{ottaviani_continuous-variable_2015, papanastasiou_finite-size_2017, zhang_finite-size_2017, lupo_parameter_2018, lupo_continuous-variable_2018, chen_composable_2018, zhao_continuous-variable_2018, ma_continuous-variable_2018} and experimental studies~\cite{liu_experimental_2019,tang_experimental_2016}.

Despite the above-mentioned advantages, the maximum transmission distance obtained from MDI-CV-QKD is limited in practical implementation~\cite{pirandola_high-rate_2015}.
Improving maximal transmission distance with a high key rate is still a challenging task.
One reason is quite low reconciliation efficiency for MDI-CV-QKD protocols with Gaussian modulation at low signal-noise-ratio~\cite{ma_long-distance_2019}, even in the presence of best error correction codes~\cite{richardson_design_2001}.
To improve the performance, some studies employed discrete modulation rather than Gaussian modulation~
\cite{silberhorn_continuous_2002, wang_continuous-variable_2019, zhao_continuous-variable_2018,ma_continuous-variable_2018, ye_enhancing_2021, leverrier_unconditional_2009, leverrier_continuous-variable_2011, huang_state-discrimination_2014}.
This discrete modulation leads to better reconciliation efficiency at a low signal-to-noise ratio, which results in enhanced transmission distance~\cite{leverrier_unconditional_2009}.
However, discrete modulation variance should be sufficiently low to derive Eve's Helvo information. This efficient low variance may cause weak low power in transmitting the quantum signal, which affects the performance of QKD.

In addition to discrete modulation, squeezed states \cite{zhang_continuous-variable_2014}, optical amplifiers~\cite{wang_continuous-variable_2019}, phase noise estimation using Bayesian inference~\cite{zhao_phase-noise_2020}, quantum scissor~\cite{jafari_discrete-modulation_2022}, and single photon subtraction operations~\cite{zhao_continuous-variable_2018, ma_continuous-variable_2018, bartley_strategies_2013,wu_improving_2015,li_non-gaussian_2016} have been included in MDI-CV-QKD protocols to improve the performance. 
Among them, single photon subtraction methods seem favorable in performance.  
However, single photon subtraction still has a low success probability (less than 0.25) of implementation at a given certain variance of EPR state~\cite{guo_continuous-variable_2019,ye_improvement_2019}. 
As a result, information between Alice and Bob can be lost during the distillation process of key rates.

Recently, quantum catalysis operations have been studied~~\cite{lvovsky_quantum-optical_2002,guo_continuous-variable_2019, ye_improvement_2019, hu_multiphoton_2016, zhou_entanglement_2018, hu_continuous-variable_2017,ye_nonclassicality_2020,ye_enhancing_2021} as an alternative solution to resolve this issue. 
Especially, Zero-Photon Catalysis (ZPC) operation is noiseless and can be implemented with existing technologies~\cite{micuda_noiseless_2012}. 
For example, quantum catalysis operations have been employed in the generation of non-classical state of light~\cite{lvovsky_quantum-optical_2002}, continuous variable entanglement~\cite{hu_continuous-variable_2017}, performance improvement of Gaussian modulated CV-QKD~\cite{ye_improvement_2019, zuo_quantum_2020}, performance improvement of four-state MDI-CV-QKD~\cite{ye_enhancing_2021}, entanglement improvement~\cite{zhou_entanglement_2018}, 
nonclassicality of two-mode non-Gaussian entangled state~\cite{ye_nonclassicality_2020}, and generation of non-Gaussian state~\cite{hu_multiphoton_2016}.
Since the eight-state quantum key distribution protocol uses an increased number of states to encode the information compared to the four-state protocol. This provides a larger optimal variance, a higher level of security, distance enhancement, and higher secret key rate.

In this paper, we present a feasible method of eight-state MDI-CV-QKD protocol with discrete modulation in the presence of ZPC operation. 
Our numerical results under the same accessible parameters show that ZPC operation on eight-state MDI-CV-QKD with discrete modulation shows advantages as compared to the four-state without ZPC~\cite{ma_long-distance_2019}, four-state with ZPC~\cite{ye_enhancing_2021}, and eight-state protocol without ZPC.
The ZPC operation on eight-state MDI-CV-QKD with discrete modulation provides a high key rate as well as increases the corresponding optimal variance for extreme asymmetric and symmetric cases. Furthermore, at a low signal-to-noise ratio, the maximum transmission distance can be increased and reconciliation efficiency can be tolerated to a lower level.

Our paper is organized as follows: Section~\ref{sec2} presents the protocol of the eight-state MDI-CV-QKD protocol in the presence of ZPC. In Sec~\ref{sec3}, the secret key rate calculations for eight-state ZPC-based MDI-CV-QKD are presented. In Sec.~\ref{sec4}, we illustrate the numerical results of our proposed scheme for extreme asymmetric and symmetric cases. Finally, we conclude our results in Sec.~\ref{sec5}.

\section{Protocol}
\label{sec2}
In general, the MDI-CV-QKD scheme can be divided into two categories, one is based on the prepare-and-measure (PM) scheme and the other is an entanglement-based (EB) scheme. The PM version can easily be applied in practical implementation while the EB version is used for security analysis~\cite{li_continuous-variable_2014}. 

In PM scheme, Alice randomly prepares eight non-orthogonal coherent states $\vert \alpha_{k}^{8}\rangle= \vert  \alpha e^{i(2k+1)\pi/4} \rangle$, where $\alpha$ is a real number and $k\in \lbrace 0,1,2,...,7\rbrace$. Alice then sends one of them to Charlie through a quantum channel. The mixture of coherent states sent to Charlie by  Alice is given by
\begin{equation}
\label{eq1}
\rho=\frac{1}{8}\sum_{k=0}^{7} \vert \alpha_{k}^{8}\rangle \langle \alpha_{k}^{8}\vert.
\end{equation}
Similarly, Bob also randomly prepares eight non-orthogonal coherent states $\vert \beta_{k}^{8}\rangle= \vert  \beta e^{i(2k+1)\pi/4} \rangle$ and sends one of them to Charlie, here, $\beta$ is a real number. After receiving these states, Charlie performs Bell's state measurement and announces his results. Alice keeps her sign of quadratures unchanged however Bob modifies his sign according to Charlie's measurements. Then Alice and Bob extract the secret key rate by using information reconciliation, parameter estimation, and privacy amplification process.

Next, we focus on the EB version of eight state MDI-CV-QKD protocol with discrete modulation in the presence of ZPC. 
The schematic diagram of the protocol is shown in Fig.~\ref{fig:1}(a). 
Alice and Bob perform the discrete modulation at the same time. We first consider discrete modulation on the Alice station.
Alice prepare two-modes squeezed state $\vert\mathrm{\Psi_{8}}\rangle_{A_{1}A_{2}}$ with two modes $A_{1}$ and $A_{2}$. Two modes entangled state prepared by Alice is given by 
\begin{eqnarray}
\label{eq2}
\vert\mathrm{\Psi_{8}}\rangle_{A_{1}A_{2}} &=&\sum_{k=0}^{7}\sqrt{\lambda_{k}} \vert \phi_{k}\rangle_{A_{1}} \vert \phi_{k}\rangle_{A_{2}} \nonumber \\
&=& \frac{1}{4}\sum_{k=0}^{7}\sqrt{\lambda_{k}} \vert \phi_{k}\rangle_{A_{1}} \vert \alpha_{k}^{8}\rangle_{A_{2}}.
\end{eqnarray}
Here, $\vert \phi_{k}\rangle_{A_{1}}$ is the non-Gaussian orthogonal state and can be represented as
\begin{equation}
\label{eq3}
\vert \phi_{k}\rangle_{A_{1}}=\frac{1}{2}\sum_{m=0}^{7}e^{\frac{i(4k+1)m\pi}{4}} \vert \phi_{m}\rangle_{A_{1}},
\end{equation}%
while the state $\vert \phi_{m}\rangle_{A_{1}}$ is expressed as
\begin{equation}
\label{eq4}
\vert \phi_{m}\rangle=\frac{e^{-\alpha^{2}/2}}{\sqrt{\lambda_{k}}}\sum_{n=0}^{\infty}\frac{\alpha^{8n+k}}{\sqrt{(8n+k)!}} \vert 8n+k\rangle.
\end{equation}%
The variances of the quadrature corresponding to the two modes $A_{1}$ and $A_{2}$ are all same ($V_{A_{1}}=V_{A_{2}}=V_{A}$), $V_{A}=1+V_{M}$, where $V_{M}=2\alpha^{2}$ represents modulation of the coherent state.

For $k\in\lbrace0,1,2,....,7\rbrace$, the normalization constant $\lambda_{k}$ are found to be~\cite{zhao_phase-noise_2020}:

\begin{widetext}
\begin{small}
\begin{eqnarray}
\label{eq5}
\lambda_{0(4)}&=&\frac{1}{4}e^{-\alpha^{2}}\bigg[\cosh\alpha^{2}+\cos\alpha^{2}\pm 2\cos\frac{\alpha^{2}}{\sqrt{2}}\cosh\frac{\alpha^{2}}{\sqrt{2}}\bigg], \nonumber\\
\lambda_{1(5)}&=&\frac{1}{4}e^{-\alpha^{2}}\bigg[\sinh\alpha^{2}+\sin\alpha^{2}\pm \sqrt{2}\bigg(\cos\frac{\alpha^{2}}{\sqrt{2}}\sinh\frac{\alpha^{2}}{\sqrt{2}}+\sin\frac{\alpha^{2}}{\sqrt{2}}\cosh\frac{\alpha^{2}}{\sqrt{2}}\bigg)\bigg], \nonumber\\
\lambda_{2(6)}&=&\frac{1}{4}e^{-\alpha^{2}}\bigg[\cosh\alpha^{2}-\cos\alpha^{2}\pm 2\sin\frac{\alpha^{2}}{\sqrt{2}}\sinh\frac{\alpha^{2}}{\sqrt{2}}\bigg], \nonumber\\
\lambda_{3(7)}&=&\frac{1}{4}e^{-\alpha^{2}}\bigg[\sinh\alpha^{2}-\sin\alpha^{2}\mp \sqrt{2}\bigg(\cos\frac{\alpha^{2}}{\sqrt{2}}\sinh\frac{\alpha^{2}}{\sqrt{2}}-\sin\frac{\alpha^{2}}{\sqrt{2}}\cosh\frac{\alpha^{2}}{\sqrt{2}}\bigg)\bigg].
\end{eqnarray}
\end{small}
\end{widetext}
The corresponding covariance matrix $\gamma_{A_{1}A_{2}}$ of the state $\vert \mathrm{\Psi_{8}}\rangle_{A_{1}A_{2}}$ is given by
\begin{equation}
\label{eq6}
\gamma_{A_{1}A_{2}}=\begin{bmatrix}
X\mathbb{I}_{2} & Z_{8}\sigma_{z} \\
Z_{8}\sigma_{z} &  Y\mathbb{I}_{2} 
\end{bmatrix},
\end{equation}
where $\mathbb{I}_{2}=\rm diag[1,1]$ and $\sigma_{z}=\rm diag[1,-1]$ are the diagonal matrices, $X=Y=V=1+2\alpha^{2}$, and $Z_{8}=2\alpha^{2}\sum_{k=0}^{7}\lambda_{k-1}^{3/2}/\sqrt{\lambda}_{k}$.

A third party named David is introduced after Alice, which controls the ZPC operation.
David employed beam splitter in the path in order to implement the ZPC operation~\cite{guo_continuous-variable_2019}
A vacuum state $\vert 0 \rangle_{D}$ in an auxiliary mode $D$ is sent to the input of a beam splitter (BS) with transmittance $T$. 
At the same time, an ideal on/off detector is placed at the output port of mod $D$, which registers the same state $\vert 0 \rangle_{D}$ with no click. 
ZPC operation implemented by BS is a probabilistic event of click and no click~\cite{nunn_heralding_2021}. 
If the detector clicks, the data should be discarded because it will herald undesired output state~\cite{nunn_heralding_2021}.
The ZPC process can be considered successful if and only if no click happens. The vacuum state $\vert 0 \rangle_{D}$ between the input and output port remains the same, in other words, the population of photon numbers cannot be changed during the catalysis process~\cite{guo_continuous-variable_2019, ye_improvement_2019}. However, it is only responsible for quantum state transformation between the modes $A_1$ and $A_2$~\cite{hu_continuous-variable_2017,guo_continuous-variable_2019}.
This ZPC process can be expressed by an equivalent operator  
\begin{equation}
\label{eq7}
\hat{O}_{0}=Tr\big[ B(T)\mathrm{\hat{\Pi}_{off}}\big]=\big(\sqrt{T}\big)^{b^{\dagger}b},
\end{equation}
where $B(T)$ is the Beam Splitter operator and can be expressed as  
\begin{equation}
\label{eq8}
B(T)=e^{(a d^{\dagger} -a^{\dagger} d) \rm ~ cos^{-1} (\sqrt{T}) },
\end{equation}
and $\mathrm{\hat{\Pi}_{off}}=\vert 0 \rangle_{D}\langle 0\vert$ is a projection operator in mode $D$.
Whereas $ a~(a^{\dagger})$, $b ~(b^{\dagger})$, and $d ~(d^{\dagger})$ are the photon annihilation (creation) operators in modes $A$, $B$, and $D$, respectively.
It is clear from Eq.~\ref{eq7} that ZPC is indeed a noiseless attenuation with an attenuation gain of $\sqrt{T}$. For any input state $\vert \alpha \rangle$, the desired heralding output state after the ZPC operation can be written as
\begin{equation}
\label{eq9}
\vert \phi \rangle_{out}=\frac{\hat{O}_{0}}{\sqrt{P_{d}}}\vert \alpha \rangle =\frac{e^{P_{d}}}{\sqrt{P_{d}}} \vert \sqrt{T} \alpha \rangle, 
\end{equation}
where $P_{d}$ represents the success probability of implementing the ZPC operation and is given by the following:
\begin{equation}
\label{eq10}
P_{d}=e^{|\alpha|^{2}(T-1)}.
\end{equation}
Thus,  Eq.~\ref{eq9} clearly shows that ZPC operation only converts the quantum state $\vert \alpha \rangle $ to $\vert \sqrt{T}\alpha \rangle $ with probability $P_{d}$.
Hence, when Alice sends a random coherent state $\vert \alpha_{k} \rangle $ and then after passing through ZPC operation, the new state becomes $\vert \tilde{\alpha}_{k} \rangle= \vert\tilde{\alpha}_{k}e^{i(2k+1)\pi/4}\rangle $, where $ \tilde{\alpha}=\sqrt{T} \alpha$. 
Therefore, when David performs the ZPC operation, it attenuates the incoming state $\vert\mathrm{\Psi_{8}}\rangle_{A_{1}A_{2}}$ in mode $A_{2}$, and one obtain the traveling state
\begin{equation}
\label{eq12}
\vert\mathrm{\Psi_{8}}\rangle_{A_{1}A_{3}} =\frac{1}{4}\sum_{k=0}^{7}\sqrt{\lambda_{k}} \vert \phi_{k}\rangle_{A_{1}} \vert \tilde{\alpha}_{k}\rangle_{A_{3}}.
\end{equation}
The corresponding covariance matrix of the traveling state $\vert\mathrm{\Psi_{8}}\rangle_{A_{1}A_{3}}$ is given by
\begin{equation}
\label{eq13}
\gamma_{A_{1}A_{3}}=\begin{bmatrix}
\tilde{X}\mathbb{I}_{2} & \tilde{Z}_{8}\sigma_{z} \\
\tilde{Z}_{8}\sigma_{z} &  \tilde{Y}\mathbb{I}_{2} 
\end{bmatrix},
\end{equation}
where $\tilde{X}=\tilde{Y}=1+2T\alpha^{2}$, $\tilde{Z}_{8}=2T\alpha^{2}\sum_{k=0}^{7}\tilde{\lambda}_{k-1}^{3/2}/\sqrt{\tilde{\lambda}}_{k}$ and $\tilde{\lambda}_{k}$ can be obtained from Eq.~\ref{eq5} by replacing $\alpha$ with $\sqrt{T}\alpha$.

At Bob station, Bob independently prepares two-mode entangled state $\vert\mathrm{\Psi_{8}}\rangle_{B_{1}B_{2}}$ with same variance as that of state $\vert\mathrm{\Psi_{8}}\rangle_{A_{1}A_{2}}$, i.e. 
$V_{A}=V_{B}=1+V_{M}$.
Alice and Bob perform the same discrete modulation.
As a result, covariance matrix $\gamma_{B_{1}B_{2}}$ of Bob's state $\vert \psi{_8}\rangle_{B_{1}B_{2}}$ is the same as $\gamma_{A_{1}A_{2}}$ of Alice's state $\vert \psi{_8}\rangle_{A_{1}A_{2}}$, because displacement operator does not effect the covariance matrix.

After that Alice and Bob hold the modes $A_{1}$ and $B_{1}$ while forwarding the modes $A_{3}$ and $B_{2}$ to Charlie through the quantum channels of lengths $L_{AC}$ (between Alice and Charlie) and $L_{BC}$ (between Bob and Charlie), respectively.
After receiving the two incoming modes $A_{3}$ and $B_{2}$, Charlie performs the Bell state measurement at the beam splitter (BS) and obtains the two output modes $C_{1}$ and $C_{2}$. 
Charlie uses homodyne detection to measure the $X$ quadrature of mode $C_{1}$ and $P$ quadrature of mode $C_{2}$. 
After measurement Charlie announce his results $\lbrace X_{C_{1}},~P_{C_{2}}\rbrace$ through public channel. 
After the announcement of results, Bob modifies the mode $B_{1}$ to mode $\tilde{ B}_{1}$ by using displacement operation $D(\beta)$, where $\beta=g\lbrace X_{C_{1}}+iP_{C_{2}}\rbrace$ and $g$ is the gain of displacement. 
Using heterodyne detection, Alice measure mode $A_{1}$ to get $\lbrace X_{A},~P_{A}\rbrace$ and Bob measure the mode $\tilde{ B}_{1}$ to get the result $\lbrace X_{B},~P_{B}\rbrace$.    
Finally, Alice and Bob get the secret key rate through post-processing. 
In post-processing, Alice and Bob implement parameter estimation, information reconciliation, and privacy amplification.

\begin{figure}[h]
   \centering
   \includegraphics[width=3.25 in]{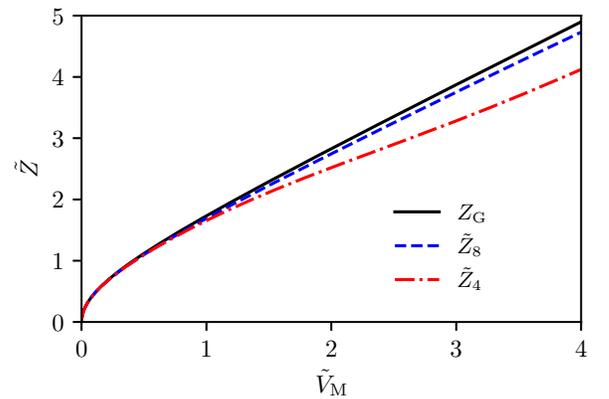}
   \caption{(Color online) The covariance factors $\tilde{Z}$ (correlation coefficient) as a function of modulation variance $\tilde{V}_{M}$ for Gaussian modulated protocol (Solid line), four-state protocol (dashed-dotted line), and eight-state protocol (dashed line).}
\label{fig2}      
\end{figure}

It may be noted that after performing the ZPC Operation, the actual modulated variance of the traveling wave $\vert \psi{_8}\rangle_{A_{1}A_{3}}$ becomes $ \tilde{V_{\rm M}}=T V_{\rm M}=T(V-1)$, with $0 < T \leq 1$.
In Fig.~\ref{fig2}, we compare the behavior of correlation coefficients of four-state discrete modulation $\tilde{Z_4}$, eight-state discrete modulation $\tilde{Z_8}$ and Gaussian modulation $Z_{G}=\sqrt{(\tilde{V}_{M}+1)^{2}-1}$.
When modulation variance $\tilde{V}_{M}<~0.5$, then $\tilde{Z_{8}}$ and $Z_{G}$ are indistinguishable in this region (along with $\tilde{Z_{4}}$). 
It means that the quantum mutual information between Bob and Eve is almost the same in both protocols. 
It is also clear from Fig.~\ref{fig2} that the amount of correlation $\tilde{Z_{8}}$ between Alice and Bob has been increased in the eight-state protocol as compared to the amount of correlation $\tilde{Z_{4}}$ in the four-state protocol. 
It implies that the secret key rate and transmission distance of the eight-state protocol is higher than the four-state protocol in the optimal regime.

\section{Secret Key Rate}
\label{sec3}

In this section, we calculate the secret key rate of ZPC-involved MDI-CV-QKD with the eight-state protocol under the Gaussian collective attack~\cite{garcia-patron_unconditional_2006, navascues_optimality_2006} where two Markovian quantum channels do not interact with each other. 
We consider that the two quantum channels i.e. Alice-Charlie and Bob-Charlie are Gaussian under the condition $V_{M} \leq 0.5$~\cite{leverrier_unconditional_2009, huang_state-discrimination_2014,shen_experimental_2010, ma_long-distance_2019}. The EB version of MDI-CV-QKD can be converted into one-way MDI-CV-QKD protocol ~\cite{ye_enhancing_2021, lupo_continuous-variable_2018, papanastasiou_finite-size_2017, chen_composable_2018,zhao_continuous-variable_2018, ma_continuous-variable_2018} under the assumption that both the state $\vert\mathrm{\Psi_{8}}\rangle_{B_{1}B_{2}}$ and displacement operation $D(\beta)$ are untrusted, except for the heterodyne detections. This equivalent one-way protocol is shown in Fig.~ 1(b). 

In~Fig.~1(a), we assume that losses of both quantum channels Alice-Charlie and Bob-Charlie are $\mu=0.2$ dB/km. The transmittance and excess noise of the quantum channels are $T_{A}~(T_{B})$ and $\epsilon_{A}(\epsilon_{B})$, respectively. The transmittance $T_{A}$ and $T_{B}$ are given as $T_{A}=10^{-\mu L_{AC}/10}$ and $T_{B}=10^{-\mu L_{BC}/10}$. 
For equivalent one-way protocol, the transmittance is $T_{C}=\frac{g^{2}T_{A}}{2}$~\cite{li_continuous-variable_2014}, and excess noise is given by~\cite{li_continuous-variable_2014}
\begin{equation}
\label{eq14}
\varepsilon_{th}=1+\chi_{A}+\frac{T_{B}}{T_{A}} \tilde{\chi},
\end{equation}
where, 
\begin{equation}
\tilde{\chi}= (\chi_{B}-1) + \bigg( \sqrt{\frac{2 (V_{B}-1)}{g^{2}T_{B}}}-\sqrt{V_{B}+1}\bigg)^{2}.
\end{equation}
Here, $\chi_{i}=(1-T_{i})/T_{i}+\epsilon_{i}$, with $i\in~\lbrace A,~B\rbrace$. Furthermore, the parameter $g$ is the gain of the displacement operation.
If we consider $g^{2}=\frac{2(V_{B}-1)}{T_{B}(V_{B}+1)}$ then the minimum value of equivalent excess noise is found to be
\begin{equation}
\label{eq15}
\varepsilon_{th}=1+\chi_{A}+\frac{T_{B}}{T_{A}}(\chi_{B}-1).
\end{equation}
 By assuming the Charlies detectors to be ideal, the total added channel noise in the shot noise limit is given by
\begin{equation}
\label{eq16}
\chi_{t}=\frac{1}{T_{C}}-1+\varepsilon_{th}.
\end{equation}
After Bell State Measurement and displacement operation, the final covariance matrix $\gamma_{A_{1}\tilde{B}_{1}}$ of the state $\vert\mathrm{\Psi_{8}}\rangle_{A_{1}\tilde{B}_{1}}$ is expressed as  
\begin{equation}
\label{eq17}
\gamma_{A_{1}\tilde{B}_{1}}=\begin{bmatrix}
a\mathbb{I}_{2} & c\sigma_{Z} \\
c\sigma_{Z} &  b\mathbb{I}_{2}
\end{bmatrix}
=\begin{bmatrix}
\tilde{X}\mathbb{I}_{2} & \sqrt{T_{c}}\tilde{Z}_{8}\sigma_{Z} \\
\sqrt{T_{c}}\tilde{Z}_{8}\sigma_{Z} &  T_{c}(\tilde{Y}+\chi_{t})\mathbb{I}_{2} 
\end{bmatrix}.
\end{equation} 
The secret key rate (SKR) of ZPC-based MDI-CV-QKD protocol with discrete modulation under the collective Gaussian attacks is modified as~\cite{leverrier_unconditional_2009,lode}
\begin{equation}
\label{eq18}
SKR=P_{d}[\beta I_{AB}-\chi_{BE}].
\end{equation}
Here, $P_{d}$ denotes the success probability given by Eq.~\ref{eq10}, $\beta$ is the reverse reconciliation efficiency, $I_{AB}$ represents the Shannon mutual information between Alice and Bob which can be written as~\cite{ho173}
\begin{equation}
\label{eq19}
I_{AB}=\log_{2}\Bigg[ \frac{a+1}{a+1-\frac{c^{2}}{b+1}} \Bigg].
\end{equation}
The coefficients $a$, $b$ and $c$ are given in Eq.~\ref{eq17}.
The term $\beta I_{AB}$ represents extracted amount of information between Alice and Bob through reverse reconciliation.
The $\chi_{BE}$ represents the Holevo bound between Bob and Eve.
The parameter $\chi_{BE}$ is bounded by correlations between Alice and Bob's data, using a Heisenberg type inequality~\cite{leverrier_unconditional_2009}.
We assume that Eve is aware of third-party David and can purify the whole system $\rho_{A_{1}\tilde{B}_{1}ED}$ to obtain the Holevo quantity between Bob and Eve as
\begin{equation}
\label{eq20}
\chi_{BE}=S(E)-S(E|B)=S(A_{1}\tilde{B}_{1})-S(A_{1}|\tilde{B}_{1}^{m_{B}}).
\end{equation}
Here, $S(A_{1}\tilde{B}_{1})=\sum_{j=1}^{2}G (\frac{\kappa_{j}-1}{2})$ is a function of the symplectic eigenvalues $\kappa_{1,2}$ of the covariance matrix $\gamma_{A_{1}\tilde{B}_{1}}$.  
The symplectic eigenvalues $\kappa_{1,2}$ are $\kappa_{1,2}=[\Delta\pm\sqrt{\Delta^{2}-4F^{2}}]/2$, where $\Delta=a^{2}+b^{2}-2c^{2}$ and $F=ab-c^{2}$.
Where as, $S(A_{1}|\tilde{B}_{1}^{m_{B}})=G(\frac{\kappa_{3}-1}{2})$ is a function of the symplectic eigenvalue $\kappa_{3}$ of the covariance matrix $\gamma_{A_{1}\tilde{B}_{1}^{m_{B}}}$. 
The covariance matrix $\gamma_{A_{1}\tilde{B}_{1}^{m_{B}}}$ can be calculated by the following relation:
\begin{equation}
\label{eq21}
\gamma_{A_{1}\tilde{B}_{1}^{m_{B}}}=a\mathbb{I}_{2}-c\sigma_{Z}(b\mathbb{I}_{2}+\mathbb{I}_{2})^{-1}c^{T}\sigma_{Z}.
\end{equation}
The corresponding symplectic eigenvalue is $\kappa_{3}=a-\frac{c^{2}}{b+1}$.
The Von Neumann entropy is given by $G(x)=(x+1)\log_{2}(x+1)-x\log_{2}(x)$. 
 

\begin{figure}[t]
\begin{tabular}{@{}cccc@{}}
\includegraphics[width=3.25 in]{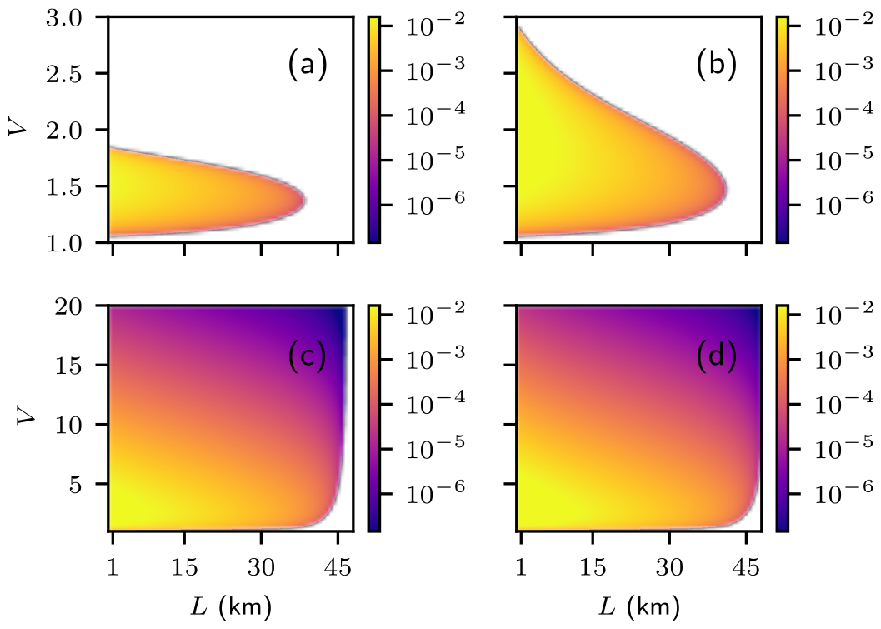}
\end{tabular}
\caption{Extreme asymmetric case: The secret key rate as a function of variance $V$ and transmission distance $L$ when optimized over transmittance $T$: (a) four-state without ZPC (b) eight-state without ZPC, (c) four-state with ZPC, (d) eight-state with ZPC. The other parameters are reconciliation efficiency $\beta=0.95$ and excess noise $\epsilon=0.002$. The value of modulation variance which leads to optimal performance is $V=1.4$ (four-state without ZPC), $V=1.5$ (eight-state without ZPC), $V=2.5$ (four-state with ZPC), and $V=2.6$ (eight-state with ZPC). It is clear that ZPC operation broadens the range of modulation variances.}
\label{fig3}
\end{figure}
\begin{figure}[t]
\centering
\includegraphics[width=3.25 in]{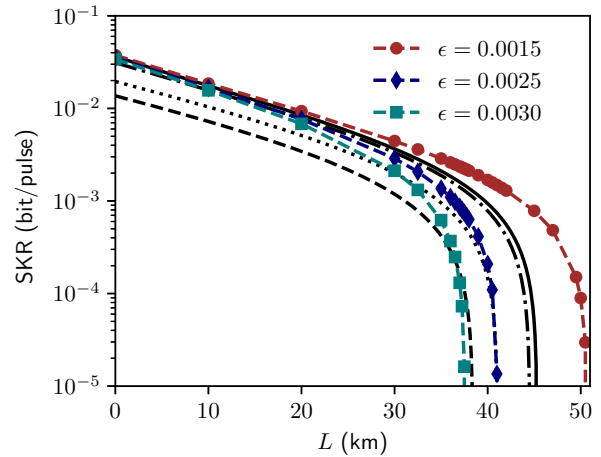}
\caption{Extreme asymmetric case: The logarithmic secret key rates as a function of transmission distance for four-state protocol without ZPC (dashed curve), eight-state protocol without ZPC (dotted curve), four-state protocol with ZPC (dashed-dotted curve), and eight-state protocol with ZPC (solid curve). The rest of the parameters are the same as in Fig.~\ref{fig3} with optimization of $T$ and optimal values of variances. 
For eight-state discrete modulated MDI-CV-QKD with ZPC, the solid, diamond, and square data points show the results at three different excess noises $\epsilon=0.0015, 0.00225$, and $0.0030$, respectively. }
\label{fig4}      
\end{figure}

\section{Performance analysis}
\label{sec4}
In this section, we analyze the performance of the eight-state MDI-CV-QKD protocol with and without ZPC operation. It has already been shown that the asymmetric case in which $L_{AC}\neq L_{BC}$ has the best performance as compared to the symmetric case $L_{AC}= L_{BC}$~\cite{pirandola_high-rate_2015,chen_composable_2018, lupo_parameter_2018}. 
In an extreme asymmetric case, Charlie is very close to Bob i.e., $L_{BC}=0$, the total transmission distance between Alice and Bob $L_{AB}=L_{AC}+L_{BC}$ has been increased under same parameters and more applicable for point-to-point communication. The symmetric case is suitable for short-range applications such as in the case of a quantum repeater.
In the following, we consider the effects of transmission distance, variance, tolerable excess noise, and reconciliation efficiency on secret key rates for both extreme asymmetric and symmetric cases.

\subsection{Extreme Asymmetric Case}
\label{sec4.1}
Modulation variance $V$ and its optimal area is an important parameter that affects the secret key rate generation of MDI-CV-QKD.
It may be mentioned that actual modulation variance in the case of the ZPC operation involved is $\tilde{V}_{M}=T(V-1)$ according to the relation $\tilde{\alpha}=\sqrt{T} \alpha$. That means the value of $V$ can be expanded by employing ZPC operation on discrete modulated MDI-CV-QKD.
To understand this point, we plot the secret key rates as a function of modulation variance $V$ and transmission distance $L=L_{AB}=L_{AC}$~(km) in the extreme asymmetric case as shown in Fig.~\ref{fig3}. 
The other parameters are reconciliation efficiency $\beta=0.95$, and excess noise $\epsilon=\epsilon_{A}=\epsilon_{B}=0.002$~\cite{jouguet_experimental_2013}.
Fig.~\ref{fig3} (a) and (b) show the results of numerical simulations for the four-state protocol without ZPC and the eight-state protocol without ZPC, respectively.
In the absence of David when no ZPC operation is involved ($T=1$), it is clear from Fig.~\ref{fig3} (a) and (b) that the optimal area of modulation variance gradually gets smaller with the increase of transmission distance. As a result, key rate generation also decreases.
It is also evident that the range of modulation variance becomes approximately 1.5 times bigger in the eight-state protocol as compared with the four-state protocol apart from the increase in transmission distance.
The secret key rate always has a peak value in the range of transmission distance, which leads to optimal performance. In other words, an optimum value of $V$ is about $1.4$ for four-state and about $1.5$ for eight-state protocol without ZPC as evident from Fig.~\ref{fig3} (a) and (b). We fixed these optimum values of $V$ for these protocols in the rest of the numerical simulations.

Fig.~\ref{fig3} (c) and (d) show the results of numerical simulations for the four-state protocol with ZPC, and the eight-state protocol with ZPC, respectively.
The secret key rate is optimized over the transmittance $T$.
When ZPC operation is involved in discrete modulated MDI-CV-QKD, the range of modulation variance $V$ broadens, and transmission distance $L$ is longer at higher key rates than without involving ZPC as shown in Fig.~\ref{fig3} (c) and (d). In other words, key rates decrease more slowly over a wider range of approximately $V=10$. 
Therefore, introducing ZPC in discrete modulated MDI-CV-QKD protocols can bring more flexible and stable performance. Transmission distance in four-state MDI-CV-QKD protocol with ZPC operation saturated at around 45~km, whereas transmission distance further enhanced to 50~km in eight-state MDI-CV-QKD protocol with ZPC operation, a longer distance than rest of the three possibilities. The optimum value of $V$ involving ZPC operation in four-state MDI-CV-QKD protocol is 2.5 while in eight-state MDI-CV-QKD protocol is 2.6 as evident from Fig.~\ref{fig3} (c) and (d).

To further understand the enhancement of transmission distances and key rates, we fixed the optimum values of variances $V$ when optimized over transmittance $T$. Figure~\ref{fig4} shows secret key rates as a function of transmission distance for four-state protocol without ZPC ($V=1.4$, dashed curve), eight-state protocol without ZPC ($V=1.5$, dotted curve), four-state protocol with ZPC ($V=2.5$, dashed-dotted curve), and eight-state protocol with ZPC ($V=2.6$, solid curve). 
The other fixed parameters are $\beta=0.95$ and $\epsilon=0.002$. 
It is clear that the secret key rate and transmission distance of the ZPC-involved discrete modulated eight-state MDI-CV-QKD system can be higher than the other three systems. 
The reason is that ZPC introduces noiseless attenuation, which increases the maximal tolerable excess noise~\cite{fiurasek_gaussian_2012}.
We also noted that tolerable excess noise $\epsilon$ significantly affects the performance of key rates and transmission distance even in the presence of ZPC operation.
As an example, in Fig.~\ref{fig4}, we consider eight-state discrete modulated MDI-CV-QKD with ZPC operation at three different excess noises $\epsilon=0.0015$, $\epsilon=0.0025$, and $\epsilon=0.0030$. 
For instance, when $\epsilon=0.0015$, maximal transmission distance further increased 5~km.
The increase of excess noise significantly shorter the transmission distance. At a longer distance, key rates fall faster. However, at a shorter distance, the effect on key rate fall is minimum. 
Therefore, extreme asymmetric ZPC-involved protocols are sensitive to excess noise at longer distances.

\begin{figure}[t]
\centering
\includegraphics[width=3.25 in]{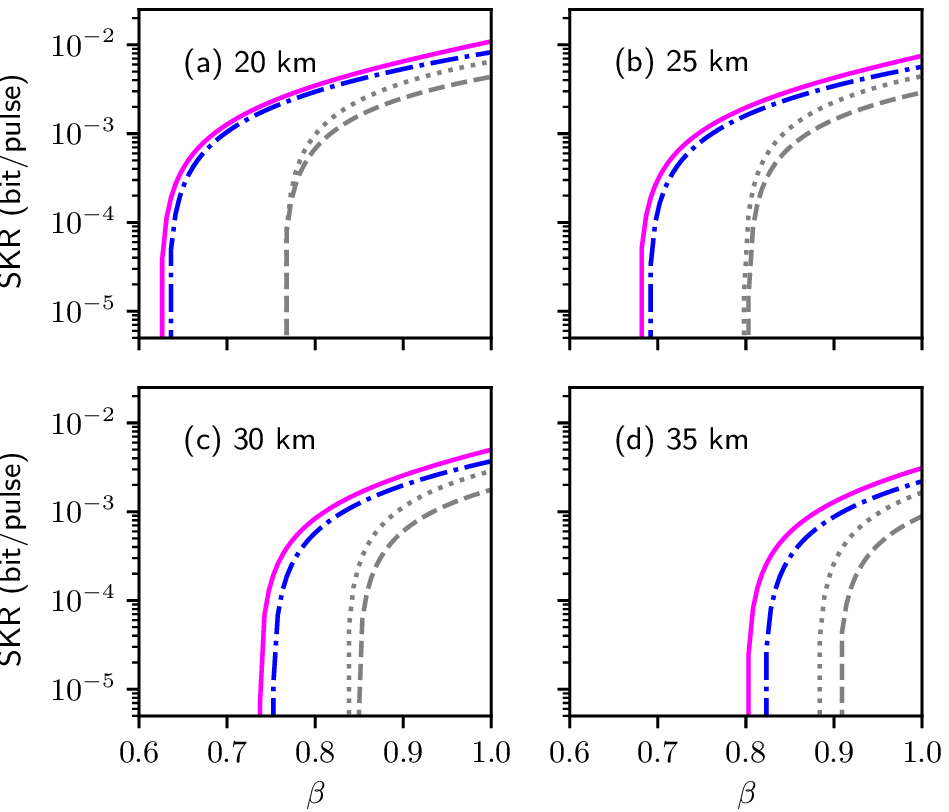}
\caption{Extreme asymmetric case: Effects of reconciliation efficiency $\beta$ on secret key rate at four different distances when optimized over transmittance $T$: (a) 20~km, (b) 25~km, (c) 30~km, (d) 35~km. The dashed, dotted, dashed-dotted, and solid curves represent the results of four-state without ZPC, eight-state without ZPC, four-state with ZPC, and eight-state with ZPC, respectively. The excess noise is fixed at $\epsilon = 0.002$ and the corresponding optimal values of variances are chosen from Fig.~\ref{fig3}.}
\label{fig5}      
\end{figure}

To extract the secret key information, the reconciliation coefficient $\beta$ is an important indicator. For the extreme asymmetric case, in Fig.~\ref{fig5}, we show the results of the secret key rate as a function of $\beta$ at four different transmission distances when optimized over $T$.
The dashed, dotted, dashed-dotted, and solid curves represent the results for four-state without ZPC, eight-state without ZPC, four-state with ZPC, and eight-state with ZPC, respectively.
The usable range of $\beta$ gets narrow with the increase of transmission distance in all four cases.
Moreover, at given transmission distances, ZPC involved an eight-state discrete modulated MDI-CV-QKD system is always better than the rest of the three protocols regarding the secret key rate.
The secret key rate falls very quickly as the reconciliation efficiency $\beta$ decreases when no ZPC operation is involved as shown by the dashed and dotted curve. 
Solid curves indicate that ZPC-involved eight-state discrete modulated MDI-CV-QKD can better tolerate the lower values of $\beta$ and also has a better key rate as compared to the four-state ZPC-involved protocol. Thus, ZPC-involved protocol with eight-state results in efficient error correction even at low SNR.

\subsection{Symmetric case}
\label{sec4.2}
Next, we consider the symmetric case in which third party Charlie is right in the middle of Alice and Bob i.e., $L_{AC}=L_{BC}$.
\begin{figure}[t]
\centering
\includegraphics[width=3.25 in]{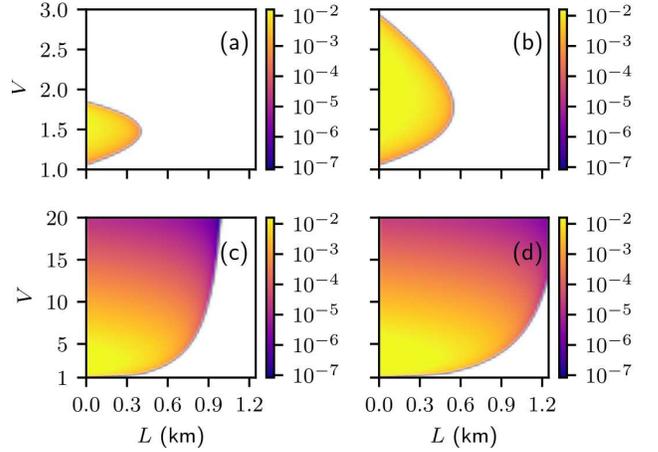}
\caption{Symmetric case: The secret key rate as a function of variance $V$ and transmission distance $L$ when optimized over transmittance $T$. (a) four-state without ZPC (b) eight-state without ZPC, (c) four-state with ZPC, (d) eight-state with ZPC. The other parameters are reconciliation efficiency $\beta=0.95$, and excess noise $\epsilon=0.002$. The value of modulation variance which leads to optimal performance is $V=1.5$ for four-state without ZPC, $V=1.8$ for eight-state without ZPC, $V=2.6$ for four-state with ZPC, $V=2.7$ for eight-state with ZPC. }
\label{fig6}      
\end{figure}
Symmetric configuration is not optimal concerning the key rate and transmission distance as compared to extreme asymmetric case.
However, the symmetric configuration has potential in short-range network applications. 
Therefore, it is still very important to analyze the performance of key rates in the symmetric case.
Similar to the extreme asymmetric case, here again, we numerically calculate the optimal value of modulation variance $V$ which results in a maximum secret key rate over the transmission distance. The results are shown in Fig.~(\ref{fig6}) when optimized over $T$ for fixed parameters $\beta=0.95$ and $\epsilon=0.002$ in ZPC-involved cases. For non-ZPC involved cases, $T=1$.
Similar to the asymmetric case, when no ZPC operation is involved, the optimal area of modulation variance gradually gets smaller with the increase of transmission distance as shown in Fig.~\ref{fig6} (a) and (b).
However, transmission distance gets significantly shorter to less than one kilometer. It is also evident that transmission distance in the eight-state protocol gets enhanced as compared to the four-state protocol.
For optimum performance,  the optimum value of $V$ is about $1.5$ for four-state and about $1.8$ for eight-state protocol without ZPC as evident from Fig.~\ref{fig6} (a) and (b). We fixed these optimum values of $V$ for the rest of the numerical simulations.

When ZPC operation is involved, the range of modulation variance $V$ over which key rate is approximately 10 times broader than without involving ZPC as shown in Fig.~\ref{fig6} (c) and (d).
In other words, key rates decrease more slowly over a wider modulation variance range similar to the extreme asymmetric case. 
Transmission distance in four-state MDI-CV-QKD protocol with ZPC operation is saturated at around 0.9~km, whereas transmission distance further enhanced to 1.2~km in eight-state MDI-CV-QKD protocol with ZPC operation, a longer distance than other three protocols.
In the whole picture, under the same parameter condition, the transmission distance for all four symmetric systems is very shorter than that of all four extreme asymmetric systems as can be observed from Fig.~\ref{fig3} and Fig.~\ref{fig6}.
The optimum value of $V$ involving ZPC operation in four-state MDI-CV-QKD protocol is 2.6 while in eight-state MDI-CV-QKD protocol is 2.7 as evident from Fig.~\ref{fig6} (c) and (d).


\begin{figure}[t]
\centering
\includegraphics[width=3.25 in]{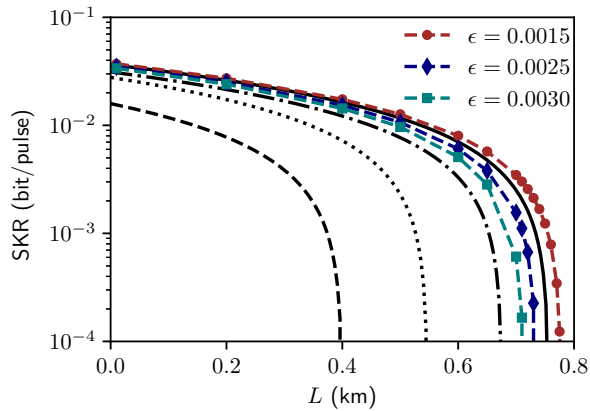}
\caption{Symmetric case: The secret key rates as a function of transmission distance for four-state protocol without ZPC (dashed curve), eight-state protocol without ZPC (dotted curve), four-state protocol with ZPC (dashed-dotted curve), and eight-state protocol with ZPC (solid curve). The rest of the parameters are the same as in Fig.~\ref{fig6} with optimization of $T$. Corresponding optimal values of variances $V$ are taken from Fig.~\ref{fig6}. 
For eight-state discrete modulated MDI-CV-QKD with ZPC, the solid, diamond, and square data points show the results at three different excess noises $\epsilon=0.0015, 0.0025$, and $0.0030$, respectively. }
\label{fig7}      
\end{figure}
\begin{figure}[t]
\centering
\includegraphics[width=3.25 in]{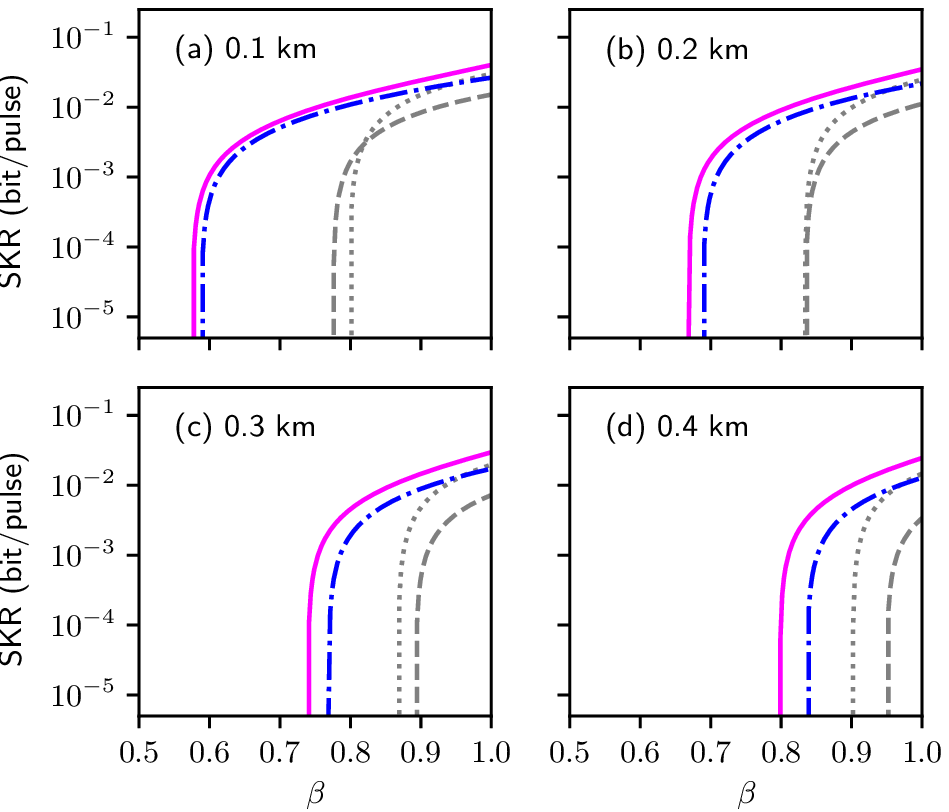}
\caption{Symmetric case: Effects of reconciliation efficiency $\beta$ on the secret key rate at four different distances when optimized over transmittance $T$: (a) 0.1~km, (b) 0.2~km, (c) 0.3~km, (d) 0.4~km. The dashed, dotted, dashed-dotted, and solid curves represent the results of four-state without ZPC, eight-state without ZPC, four-state with ZPC, and eight-state with ZPC, respectively. The excess noise is fixed at $\epsilon = 0.002$ and the corresponding optimal values of variances are chosen from Fig.~\ref{fig6}.}
\label{fig8}      
\end{figure}

\begin{figure}[t]
\centering
\includegraphics[width=3.25 in]{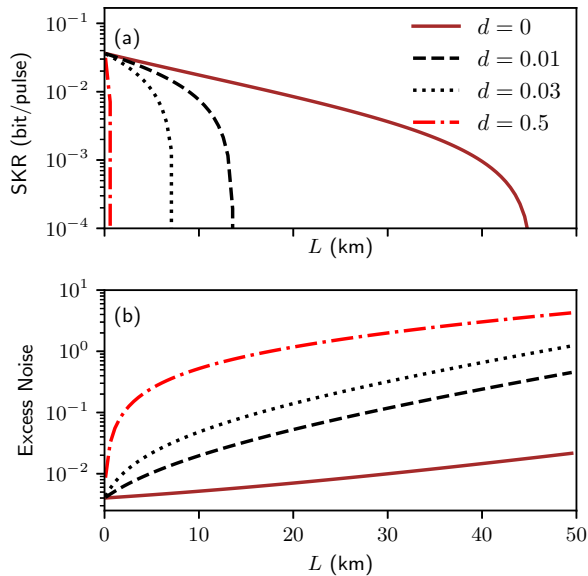}
\caption{(a) Effects of transition from extreme asymmetric to symmetric configuration on secret key rate versus transmission distance. The fixed parameters are $\beta=0.95$, excess noise $\epsilon=0.002$, and $V=2.7$.Here, $d$ represent ratio of $L_{BC}$ to $L_{AC}$. 
(b) The corresponding total equivalent excess noise as a function of transmission distance.}
\label{fig9}      
\end{figure}


Next, we fixed the optimum values of variances $V$ when optimized over transmittance $T$ and calculate the secret key rate as a function of transmission distance, which is shown in Fig.~\ref{fig7}. In Fig.~\ref{fig7}, the dashed curve shows the four-state protocol without ZPC ($V=1.5$), the dotted curve eight-state protocol without ZPC ($V=1.8$), the dashed-dotted curve shows the four-state protocol with ZPC ($V=2.6$), and 
solid curve shows eight-state protocol with ZPC ($V=2.7$), for fixed parameters $\beta=0.95$ and $\epsilon=0.002$.
It is clear that the secret key rate and transmission distance of the ZPC involved discrete modulated eight-state MDI-CV-QKD system is also enhanced in the symmetric case like extreme asymmetric case.
In order to see the effect of tolerable excess noise $\epsilon$ on key rate and transmission distance on eight-state discrete modulated MDI-CV-QKD with ZPC operation, we plot the result for three different choices of excess noises $\epsilon=0.0015$ (circles), $\epsilon=0.0025$ (diamonds), and $\epsilon=0.0030$ (squares) in Fig.~\ref{fig7}.
 The increase of excess noise shortened the transmission distance in the symmetric case.

Next, we show the numerical results of the secret key rate as a function of reconciliation coefficient $\beta$ at four different transmission distances when optimized over $T$ in Fig.~\ref{fig8}. 
The dashed, dotted, dashed-dotted, and solid curves represent the results for four-state without ZPC, eight-state without ZPC, four-state with ZPC, and eight-state with ZPC, respectively.
At given transmission distances, the ZPC-involved eight-state discrete modulated MDI-CV-QKD system is better than the rest of the three protocols with respect to the secret key rate and tolerable excess noise like extreme asymmetric case.

Finally, in Fig~\ref{fig9} (a), we present the numerical results showing the transition from extreme asymmetric to the symmetric case. 
In order to do this, we define $L_{BC}=d \times L_{AC}$, and $L=(1-d) \times L_{AC}$, where $d$ represent ratio of $L_{BC}$ to $L_{AC}$. Clearly when the distance between Charlie and Bob increases transmission distance dramatically decreases. This dramatic decrease is because of sensitivity to excess noise~\cite{ma_long-distance_2019}. In order to see the effect of excess noise, we calculate total equivalent excess noise from Eq.~\ref{eq15} for both extreme asymmetric ($d=0$) and symmetric cases ($d \neq 0$). It can be clearly seen from Fig~\ref{fig9} (b) that the gap between excess noise of symmetric and extreme asymmetric cases becomes larger and larger when transmission distance increases. Since transmission distance is sensitive to excess noise as observed in numerical results of Fig~\ref{fig4} and Fig~\ref{fig7}, therefore, the symmetric case leads to poorer performance as compared to the extreme asymmetric case.

\section{Conclusion}
\label{sec5}

We have suggested a performance improvement of the MDI-CV-QKD with discrete modulation by performing the ZPC operation on an eight-state protocol. 
This eight-state-based scheme may offer a possibility to further extend the maximal transmission distance. 
For extreme asymmetric and symmetric cases in an asymptotic regime, our results show that the secret key rate of the ZPC-involved discrete modulated MDI-CV-QKD protocol can be increased, as compared with the original protocol eight-state protocol, original four-state protocol, and original four-state protocol with ZPC.
 Furthermore, we find that our eight-state protocol enables the discrete modulated MDI-CV-QKD system to tolerate a lower reconciliation efficiency. 
 Furthermore, it is observed that the performance of the extreme symmetric case is much better than the symmetric case due to the sensitivity of excess noise.
 
It is important to note that a successful ZPC operation requires heralding signals based on the detection of zero photons (no click event), which forms the basis of noiseless attenuation for quantum communications~\cite{micuda_noiseless_2012}. 
The experimental detection of zero photons presents several challenges such as distinguishing zero photon and dark counts on detectors.
However, combining the detection of a reference (timing) signal with the simultaneous absence of detection in a single photon detector can successfully register a “no-click” as recently demonstrated in an interesting experiment by C.M. Nunn \textit{et al}.,~\cite{nunn_heralding_2021}. Furthermore, this basic approach reverses the problematic roles of detector inefficiency and dark counts in realistic detectors.

\begin{acknowledgments}
The authors SQ, MW, and MI acknowledge the support of the HEC NRPU Research project No 15313.
\end{acknowledgments}

 \section*{Conflict of interest}
 The authors declare that they have no conflict of interest.

 \section*{Data availability}
Data underlying the results presented in this paper may be obtained from the authors upon reasonable request.
\bibliographystyle{apsrev4-2}
\bibliography{QKD-BILAL.bib}



\end{document}